\documentclass[aps,twocolumn,groupedaddress]{revtex4}

\usepackage{graphicx}

\begin{document}

\title{Reformation of an oblique shock observed by Cluster}
\thanks{An edited version of this paper was published by AGU.  
Copyright (2009) American Geophysical Union. 
Lefebvre, B., Y. Seki, S. J. Schwartz, C. Mazelle (2009),
{\it Reformation of an oblique shock observed by Cluster},
J. Geophys. Res. {\bf 114}, A11107, doi:10.1029/2009JA014268. 
{\tt http://dx.doi.org/10.1029/2009JA014268}}

\author{Bertrand Lefebvre}
\affiliation{
Space Science Center,
University of New Hampshire,
Durham, NH 03824, USA
}
\homepage{http://www.spaceplasma.unh.edu/~bertrand}

\author{Yoshitaka Seki}
\affiliation{
Institute of Space and Astronomical Science, Japan Aerospace Exploration Agency,
Sagamihara, Japan.
}

\author{Steven J. Schwartz}
\affiliation{
Space and Atmospheric Physics,
The Blackett Laboratory,
Imperial College London, London SW7 2BW, UK
}

\author{Christian Mazelle}
\affiliation{
CESR / UPS-CNRS, 9 Avenue du Colonel Roche, Toulouse, 31400, France
}

\author{Elizabeth A. Lucek}
\affiliation{
Space and Atmospheric Physics,
The Blackett Laboratory,
Imperial College London, London SW7 2AZ, UK
}

\begin{abstract}
On 16 March 2005, the Cluster spacecraft crossed a shock almost at the transition between the quasi-perpendicular and quasi-parallel regimes ($\theta_{Bn}=46^{\circ}$) 
preceded by an upstream low-frequency ($\approx$ 0.02 Hz in the spacecraft frame) wavetrain observed for more than 10 mn. The wave semi-cycle nearest to the shock was found to grow in time, steepen and reflect an increasing fraction of the incoming ions. This gives strong indication that this pulsation is becoming a new shock front, standing $\sim 5\lambda_p$ upstream of the main front and growing to shock-like amplitude on a time-scale of $\sim 35\Omega_p$. Downstream of the main shock transition, remnants of an older front are found indicating that the reformation is cyclic. This provides a unique example where the dynamics of shock reformation can be sequentially followed. The process shares many characteristics with simulations of reforming quasi-parallel shocks.
\end{abstract}

\maketitle

\section{Introduction}

The question of the stability of the shock structure has regained recent interest with the advent of multipoint measurements at the Earth's bow shock. Numerical simulations have indicated for a long time that even under steady upstream conditions and for a broad range of parameters collisionless shocks can exhibit dramatic structural changes and eventually self-reform quasi-periodically \citep{BW72JGR,Bur89GRL,LGS04SSR}. However experimental evidence of this kind of behaviour in space plasmas have been limited. For instance, \citet{TGB90JGRa} and \citet{TGO93JGR} have observed quasi-periodic variations on a time-scale of approximately two upstream proton gyroperiods in the ion velocity distributions downstream of the quasi-parallel Earth's bow shock which they have attributed to shock reformation. Using multipoint Cluster measurements, \citet{LHD08JGR} have studied the scales and growth of large amplitude pulsations (Short Large Amplitude Magnetic Structures, SLAMS) thought to be part of the quasi-parallel shock transition region. At the quasi-perpendicular shock, \citet{HCL01AG} have noticed on some Cluster crossings a significant variability in magnetic field profiles, particularly in the foot, despite a relatively small spacecraft separation. A large variability in magnetic and electric field time series was also observed by \citet{LKB07GRL} at a high Mach number quasi-perpendicular shock crossing, complemented with bursty variations in reflected ions occuring on a time scale comparable to a proton gyroperiod. Another form of non-stationarity was found by \citet{MBH06JGR} who have identified coherent oscillations with a wavelength of a few tens of upstream ion inertial length confined to the shock layer and propagating along it.

The general problem of the shock stationarity is complicated by the importance of the shock geometry. Quasi-parallel and quasi-perpendicular shocks are indeed distinguished by significantly different structure, scales and dissipative processes
(e.g. \citep{Sch06SSR}). At least for supercritical shocks resistivity is insufficient to provide the required dissipation, which is then provided in a first step by reflecting upstream part of incoming ions. Assuming a specular reflection process and a step-like shock transition one finds out that the limit between quasi-perpendicular and quasi-parallel shocks $\theta_{Bn}=45^{\circ}$ corresponds to the angle below which the guiding-centre of the reflected ions is directed upstream instead of being directed back towards the shock \citep{GTB82GRL}. However specularly reflected ions at shocks only slightly below $\theta_{Bn}=45^{\circ}$ may still re-encounter the shock front during the course of their gyromotion unless the shock angle is actually lower than $\sim 40^{\circ}$ \citep{STG83JGR}. Only then are these reflected ions able to escape upstream, and by backstreaming against the solar wind flow they excite upstream waves (e.g. \citep{Que88JGR}). As these low-frequency waves grow to large amplitudes they interact with the incoming solar wind, giving quasi-parallel shocks a more extended and visibly more complex and dynamic structure than quasi-perpendicular ones. Clearly there is more to collisionless shock dissipation than simple specular reflection. In the case of curved shocks such as the Earth's bow shock this picture is further complicated by ions streaming from other parts of the shock which also contribute to instabilities near the quasi-parallel shock. \citet{BLS05SSR} provide a recent review of observations of quasi-parallel shock structure and processes (see also \citep{Bur95ASR}).

Observations have therefore shown that the bow shock can significantly deviate from the textbook picture of a locally planar and stationary structure and that it may eventually reform. However the scales and complexity of shock reformation combined with the limited number of measurement points have not provided so far direct evidence by following sequentially the reformation of a shock front.

This paper presents a case study of a crossing by the Cluster fleet \citep{EFG01AG} of a shock with $\theta_{Bn}\approx 45^{\circ}$. The region upstream of the shock exhibits a low-frequency quasi-monochromatic wave as more commonly observed upstream of quasi-parallel shocks. The wave cycle nearest to the shock ramp is found to steepen and grow into a pulse-like structure which we argue corresponds to the formation of a new shock-ramp.  The four Cluster spacecraft were able to sequentially observe this process which we interpret to be shock reformation.

\section{Data}

This study concentrates on a shock crossing on 16 March 2005 around 1530 UT by Cluster occurring during a data burst mode interval. The quartet was in a tetrahedron configuration with inter-spacecraft separations of the order of 1300 km.
The magnetic field data from the Flux-Gate Magnetometer (FGM) \citep{BCA01AG}  used in this study was averaged at a resolution of one vector per spin-period (4 s) or 10 vectors per second, averaged from a 67 vectors/second sampling frequency in burst mode. The electric field data from the Electric Field and Wave (EFW) \citep{GBH97SSR} instrument used here has a temporal resolution of approximately 2 ms. Ion data was provided by the Hot Ion Analyzer (HIA) of the Cluster Ion Spectrometer (CIS) \citep{RAB01AG} which measures fluxes of positive ions irrespective of species in the energy range 0.005-26 keV/e and takes a spin period to build a full 3d distribution (transmitted to the ground every spin in burst mode) and has an angular resolution of $22.5^{\circ}\times 22.5^{\circ}$. However during this interval the instrument was in a magnetospheric mode not optimized for the solar wind, and CIS/HIA data were available on spacecraft 1 and 3 only. Electron data came from the Low Energy Electrostatic Analyzer (LEEA) of the Plasma Electrons And Current Experiment (PEACE) \citep{JAB97SSR}. In this burst mode, a 3d distribution with reduced polar resolution (3DX1) was transmitted to the ground every spin, and consists of 26 energy levels in the range 0.007-1.7 keV, 6 polar and 32 azimuthal angular bins. This data was then reduced on the ground to pitch-angle distributions using high-resolution FGM data, and corrected for the spacecraft potential effect using spin-resolution EFW data. Magnetic field and particle from the MAG and SWEPAM instruments onboard the Advanced Composition Explorer (ACE) spacecraft \citep{SFM98SSR} were used to compute the upstream solar wind parameters. The time delay taken by the solar wind to travel from the Lagrange point to the Earth's bow shock was taken into account assuming it moves with a constant velocity parallel to the GSE $x$-axis.

\section{Shock observations}

\subsection{Global shock properties and Cluster configuration}

\begin{table*}[ht]
\caption{
  Main shock parameters at the asymptotic upstream and downstream locations. Upstream parameters were taken from ACE data.
  Frame-dependent quantities are given in the Normal Incidence frame (the shock rest frame where the upstream 
  solar wind velocity is directed along the shock normal). 
}
\begin{tabular}{llll}
    \tableline
    Parameters                                               & units          & upstream                   & downstream               \\  \hline
    $B$                                                      & nT & 9     & 27   \\ 
    $n_p$                                                    & cm$^{-3}$      & 5.6                      & 19.6     \\
    Proton ram energy                   & eV             & 946                    & 78                    \\ 
    $T_p$                                                    & eV             & 4                      & 198                   \\ 
    $\beta_p$                                                & --             & 0.1                      & 2.2                     \\ 

    $T_e$                                                    & eV             & 17                     & 54                    \\ 
    $\beta_e$                                                & --             & 0.5                      & 0.6                     \\  
    Proton gyrofrequency, $f_{gp}$                           & Hz             & 0.14                      & 0.41                     \\ 
    Proton inertial length, $\lambda_{p}$                   & km             & 96                     & 51                    \\ 
    Thermal proton gyroradius, $\rho_{p}$                    & km             & 22                     & 54                    \\ 
    Convected proton gyroradius, $v_{p1}/\Omega_{gp}$        & km             & 493                    & 167                   \\ 
    Specularly-reflected proton gyroradius                   & km             & 4536                   & --                       \\ 
    Phase-standing whistler wavelength, $\lambda_w$          & km             & 75                     & --                       \\ 
    $\theta_{Bn}$                                            & deg            & 47                     & 77                    \\ 
    Alfv\'en velocity, $c_A$                                 & km/s           & 77                     & 122                   \\ 
    Alfv\'en Mach number, $M_A$                              & --             & 5.5                      & 1.00                     \\ 
    Magnetosonic Mach number, $M_{ms}$                       & --             & 4.6                      & 0.8                     \\ 
    \\ \tableline
\end{tabular}\label{tab:shock_params}
\end{table*}

\begin{figure}
  \includegraphics[width=18pc]{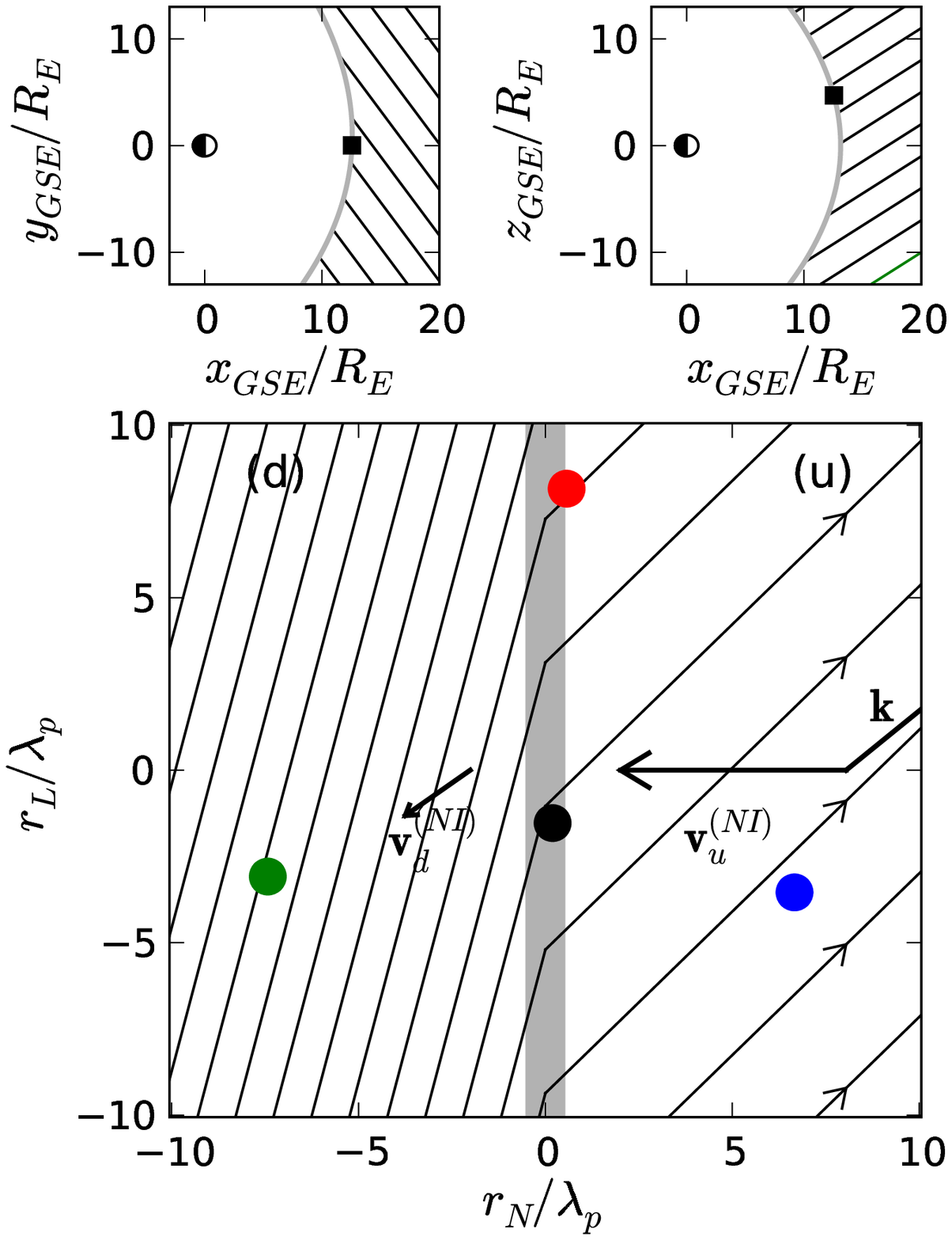}
  \caption{Shock crossing configuration. Top panels show the upstream magnetic field lines and the location of the crossing (black square) in the GSE $xy$ and $xz$ planes. The main panels show the asymptotic magnetic fields, solar wind velocity in the normal incidence shock frame, and projection onto the coplanarity (NL) plane of spacecraft locations and direction of the upstream wavector. The grey area corresponds to the main shock ramp, and distances are normalized to the upstream proton inertial length $\lambda_p$. Spacecraft in this plane are approximately aligned along the shock normal for C1 (black), 3 (green) and 4 (blue), while C2 (red) crosses the shock nearly simultaneously to C1 but approximately $10 \lambda_{p}$ away along the shock front. The spacecraft travel from downstream to upstream.}
  \label{fig:config}
\end{figure}

The solar wind as monitored by ACE, time-shifted by the solar wind travel time from the Lagrange point to the shock, remains steady and quiet during the whole time interval and for nearly half an hour before, in particular with a velocity around 400 km/s and no important change in magnetic field direction. Because of the presence of large amplitude fluctuations extending far upstream of the shock (and of the CIS instrument mode), the asymptotic upstream field and particle parameters are estimated from the ACE measurements (taking into account the solar wind travel time to the shock).

The crossing occurs at $(12.7, 0, 4.6)Re$ (GSE), from downstream to upstream (outbound crossing). The shock timing analysis (which assume a planar surface moving at constant speed) \citep{Sch00Book} yields a normal $\hat{\mathbf{n}} = (0.93, -0.12, 0.35)$ (GSE), a shock angle between the upstream magnetic field and the normal $\theta_{Bn} = 45^{\circ}$ and a velocity along normal of -13 km/s in the spacecraft frame. The Abraham-Shauner method \citep{Sch00Book} which makes use of the MHD jump (Rankine-Hugoniot) conditions for the magnetic and velocity fields (and assumes as well a planar surface and shock stationarity) yields a similar result, $\hat{\mathbf{n}} = (0.93, -0.10,  0.35)$ and $\theta_{Bn} = 46.5^{\circ}$. The shock is therefore within experimental errors at the formal limit of quasi-perpendicular and quasi-parallel shocks, and we shall generically call it an oblique shock.

The projected spacecraft locations onto the coplanarity plane show that Cluster-1 (C1), 3 and 4 are approximately aligned along the shock normal while C1 and 2 are perpendicular to it and cross the ramp nearly simultaneously (fig.\ \ref{fig:config}). The shock is crossed first by C4, then by C1 and 2 together and finally by C3.

The shock is super-critical with an Alfv{\'e}n Mach number $M_A=5.5$. The proton thermal to magnetic pressure ratio for this shock is low, $\beta_p=0.1$. Other shock and plasma parameters are summarized in Table \ref{tab:shock_params}.

\begin{figure}
  \includegraphics[width=18pc]{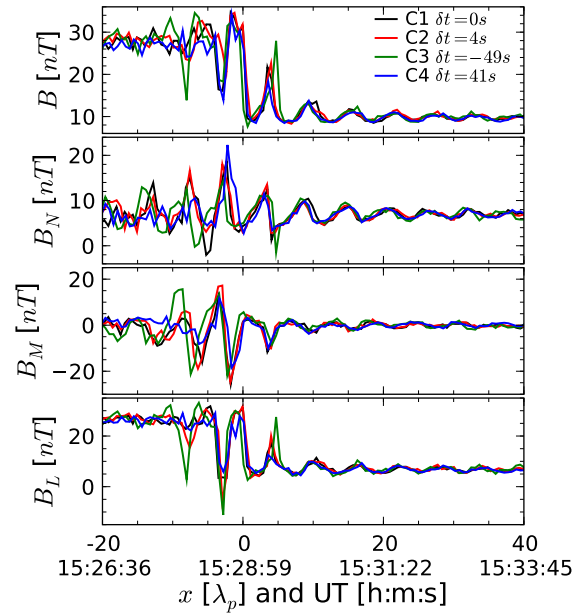}
  \caption{4s-resolution magnetic field intensity (top panel) and components in shock normal coordinates on all four spacecraft. Data are time-shifted to allow comparisons of the shock structure and upstream wavetrain measured by the spacecraft. Time information is translated into distance assuming the whole structure travels at the constant shock velocity estimated from the timings.}
  \label{fig:lowresshock}
\end{figure}

\subsection{The upstream low-frequency wave}

An upstream ultra-low frequency wave (ULF) with a period $\sim$42 s ($f=24$ mHz) in the spacecraft frame is observed for over 10 minutes from the ramp (fig.\ \ref{fig:lowresshock}). The time-series appear quasi-monochromatic at 4 s-resolution, but considerable broadband higher-frequency fluctuations are seen in the higher-resolution data. The ULF oscillations can be seen in the magnetic field intensity as well as all three components and strongly increase in amplitude in the vicinity of the ramp. For instance, at the shock ramp the magnetic field fluctuations along the shock normal reach $\delta B_N / B_N\approx 1.7$, and the fluctuations in the non-coplanar component of the magnetic field $\delta B_M$ are nearly as large as the overall difference between upstream and downstream magnetic field component in the maximum variance direction $B_L$. 
The general shock profile, ULF wave characteristics and its relative location with respect to the shock seem remarkably similar on all four spacecraft despite the large spacecraft separation and temporal spread of the crossings (the first and last shock crossings are approximately 80 s apart). This gives to the wave an unexpected appearance of phase-stationarity with respect to the shock (to be discussed in the final section). Oscillations extend over one or two periods downstream of the ramp with a shorter period $\sim$36 s, although it is not entirely clear whether they correspond to a transmitted wave or to the overshoot-undershoot cycle.

Minimum Variance Analysis (MVA, see e.g. \citep{SR09SSR}) applied suggests that the wavevector is roughly aligned with the magnetic field (the deviation, $\theta_{kB}\approx 24^{\circ}$, is mainly in the out of coplanarity plane direction), 
$\mathbf{k}/k = \pm (0.63, 0.38, 0.67)$ (NML). 
The wave has a high degree of polarization (0.92), is left-hand polarized (with respect to the magnetic field) in the spacecraft frame with ellipticity $\varepsilon = -0.81$ and has a low compression ratio $C_e=\left(\delta n/n_0\right)^2(B_0^2/\delta B^2)=0.12$ (tab. \ref{tab:wave_params}).

More properties can be derived from the time delays between spacecraft assuming a planar wavefront. Although in general for a monochromatic wave the time delays can only be determined modulo the wave period, we find that the smallest delays determined from the time-series provide a direction close to that of the MVA, namely 
$\mathbf{k}/k = (0.66,  0.37,  0.65)$ in NML coordinates
(2$^{\circ}$ away from the MVA result). The corresponding phase velocity in the spacecraft frame is 142 km/s which yields a wavelength of 5930 km. From the wavevector and the mean upstream plasma velocity, the frequency in the plasma frame is found to be 0.01 Hz and the plasma frame phase velocity is 56 km/s, lower than but comparable to the upstream Alfv\'en speed (76 km/s). In this frame, the wave propagates against the solar wind flow and therefore is right-hand polarized. The properties summarized in table \ref{tab:wave_params} such as wavelength $\sim R_E$, upstream propagation direction and right-hand polarization are fairly typical of ULF waves studied using ISEE \citep{HR83JGR} or Cluster \citep{EBD02GRL,AHL05JGR} datasets, and are thought to result from an electromagnetic ion/ion right-hand resonant instability due to ions backstreaming from the shock. Finally, the phase velocity along the shock normal in any shock rest frame is estimated to be $\mathbf{v}_{\varphi}^{\mathrm{(shock)}}\cdot\hat{\mathbf{n}}\approx -80$ km/s showing that as expected the wave is not phase-standing but convected towards the shock by the solar wind.

\begin{figure}
  \includegraphics[width=18pc]{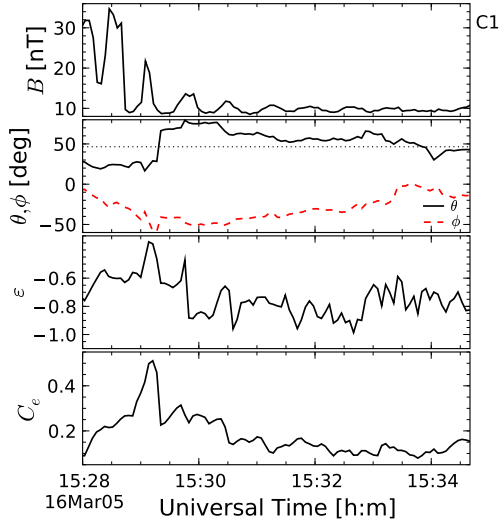}
  \caption{Wave properties in spacecraft frame derived from MVA on 84s intervals of 4s resolution data. The top panel shows the magnetic field intensity. Next panel shows the polar angle $\theta$ between $\mathbf{k}$ and $\hat{\mathbf{n}}$, and $\phi$ which is the azimuthal angle in the LM plane with respect to the L axis. The dotted lign indicates $\theta_{Bn}$. Lower panels show the ellipticity and electron compression ratio.}
  \label{fig:running_mva}
\end{figure}

Applying MVA to shorter time intervals (2 wave periods) however reveals that these properties change closer to the shock front. As shown in fig.\ \ref{fig:running_mva} the wavevector approximately aligns itself with the shock normal, the polarization becomes less circular and the compression ratio increases. These changes appear during intervals which do not yet include the shock ramp, corresponding to specific properties of the cycle nearest to the ramp which may be affected by the shock foot and reflected ions. Since the wave is assumed to be planar and has a high degree of polarization, an effect of the alignment of the wavevector with the shock normal should be to diminish the perturbations due to the wave to the planarity of the shock surface. In addition the oscillations seem to remain in-phase with the shock as if the ramp was part of one cycle and other cycles were phase-standing next to it. Indeed, the timings of the pulse nearest to the shock ramp confirm the wavevector alignment with the normal (consequently $\theta_{kB}\approx 40^{\circ}$) and a velocity nearly identical to that of the shock, about -13 km/s in the spacecraft frame. This yields a wavelength an order of magnitude lower than further upstream, $\lambda\approx 500 km \approx 5 \lambda_p$. 

Besides changes in wavevector and velocity, the wave experiences a strong amplification near the shock. The cycle standing nearest to the ramp indeed nearly reaches shock-like amplitudes and displays an interesting behaviour detailed in the next section.

\begin{table*}[ht]
\caption{Properties of the upstream ultra-low frequency wave in spacecraft and plasma frames.}
\begin{tabular}{lll}
    \tableline
              & Spacecraft & Plasma \\
    \tableline
    Frequency $/f_{gp}$ & 0.14  &  0.07  \\
    Wavelength $/\lambda_p$ & 62 & 62 \\
    Phase velocity $/c_A$ &  1.8 & 0.74 \\
    Polarization degree & 0.92 & 0.92 \\
    Polarization & Left-hand & Right-hand \\
    Ellipticity & -0.81 & + 0.81 \\
    Compression ratio & 0.12 & 0.12 \\
    \tableline
\end{tabular}\label{tab:wave_params}
\end{table*}

\subsection{The growing and steepening pulse upstream of the shock ramp}

\begin{figure}
  \includegraphics[width=19pc]{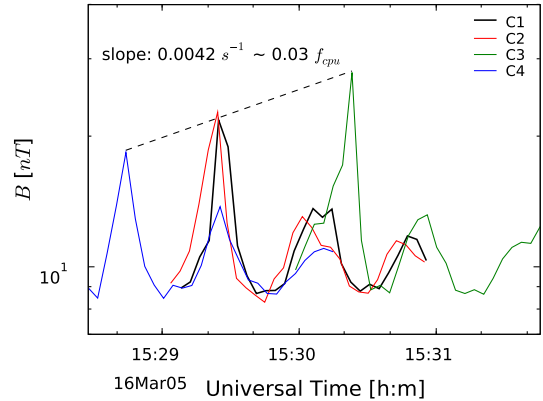}
  \caption{Main shock ramp and upstream pulse, shown with 4 s resolution FGM data. The pulse growth rate is approximately $0.03f_{cpu}$.}
  \label{fig:pulse_growth}
\end{figure}

As noted in the previous paragraph, upstream magnetic field fluctuations reach their largest amplitude at the wave cycle nearest to the shock front. This large amplitude pulse-like structure is crossed by the spacecraft in the same order as the shock ramp (C4 first, followed nearly simultaneously by C1 and C2 and then C3), showing that it is not a partial recrossing of the shock front but a distinct upstream structure. The time delay between the crossings of the shock ramp and the feature is nearly the same for all four spacecraft (about 20s), suggesting that this structure extends parallel to the shock ramp and remains at a constant distance from it.

As shown in fig.\ \ref{fig:pulse_growth} the magnetic field intensity of the structure is the lowest at the crossing by C4, larger at C1 and C2 (and slightly more so at C2 than C1) and largest at C3 where its amplitude is comparable to that of shock itself. The same observation applies to $B_L$. The structure is therefore growing in time. An exponential fit to the peak amplitudes yields a growth rate of $\gamma\approx 4\cdot 10^{-3}$ Hz $\approx 0.03 f_{gpu}$. The amplitudes are equally well fit by a linear curve with slope $0.011B_u$ s$^{-1}$ $\approx 0.08B_uf_{gpu}$. Both models estimate that it takes up to $\sim 35$ upstream proton gyroperiods for the pulse to grow to shock-like amplitudes.

\begin{figure}
  \includegraphics[width=19pc]{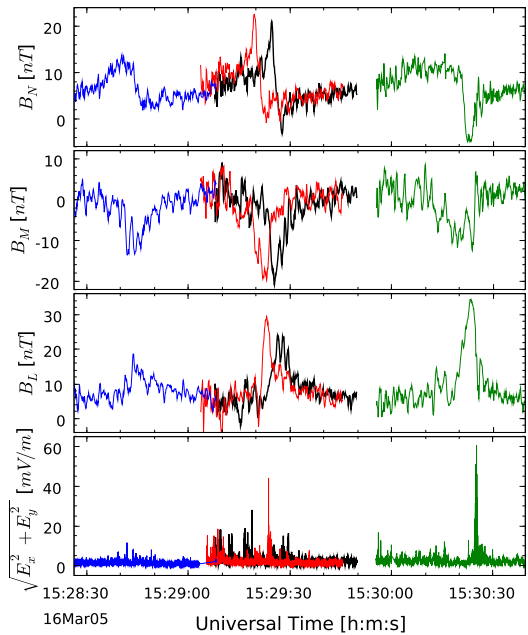}
  \caption{High-resolution (10 vectors per-second) magnetic field measurements of the growing structure in shock NML coordinates. The structure is not only found to grow in amplitude but also to steepen while emitting whistlers, and measured electric field intensities correspondingly increase to reach ramp-like values.}
  \label{fig:bump_highres}
\end{figure}

High-resolution magnetic field data show that besides growing in amplitude the structure is found to steepen (fig.\ \ref{fig:bump_highres}). The steepening is most visible in $B_L$ and occurs on the upstream edge of the pulse (apart perhaps from C2 on which the "downstream" side of the pulse seems at least as steep as its upstream side). As the steepening proceeds quasi-periodic whistler-like fluctuations (at $\approx 0.15$ Hz, best seen on $B_M$) are emitted upstream and the measured electric field magnitude $(E_x^2+E_y^2)^{1/2}$ increases too and reaches on C3 values comparable to those in the ramp. Furthermore, high-resolution data reveal that the pulse is seen slightly earlier on C2, consistent with its position slightly upstream of C1 as sketched in Fig 1. However, that data also shows the pulse's growth and steepening to be more advanced at C2. This shows that at this separation scale the pulse's structure and growth is not perfectly homogeneous along the shock plane.

Finally one notes on C1 and C2 in between the pulse and the shock ramp localized dips in $B_L$ reaching negative values, similar to that observed  during the reformation cycle in 1d \citep{Bur89GRL,WTO90JGR} and 2d \citep{SFK93JGR} quasi-parallel shock hybrid simulations.

\subsection{Particles and cross-shock potential}

\begin{figure}
  \includegraphics[width=19pc]{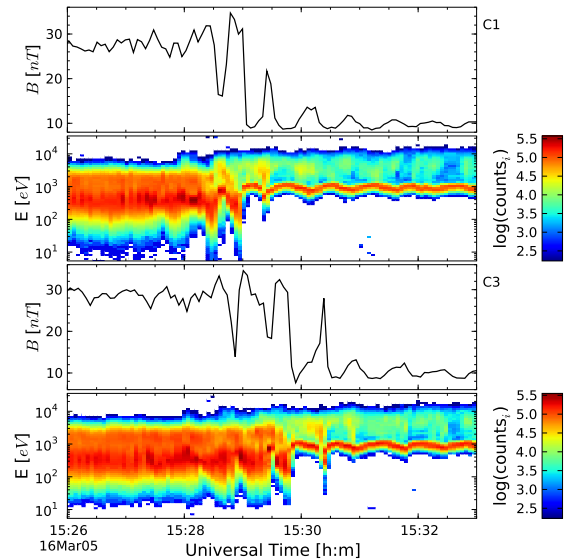}
  \caption{Energy spectrograms of ions from CIS/HIA, summed over all angles. Upstream, both solar wind and energetic ions are strongly affected by the wave. At the shock, a distinct population of reflected ions around 3 keV is observed at the shock ramp on C1 but not on C3, where it is observed on the upstream steepened structure instead.}
  \label{fig:hia_spectro}
\end{figure}

If the growing feature is indeed becoming a new shock front, then it should affect the heating and reflection of incoming solar wind particles and correspondingly develop an electric potential jump. fig.\ \ref{fig:hia_spectro} shows magnetic intensity profile and ion energy spectra (from CIS/HIA, all species and summed over all view angles) on C1 and C3. Two distinct ion populations are clearly observed in the solar wind. The lowest energy one ($\approx 1$ keV) is the incoming solar wind beam which undulates under the effect of the wave, yielding large oscillations of the velocity moment. It seems however that the wave results in little or no ion heating (which is nevertheless difficult to check quantitatively in the absence of reliable temperature estimates due to the particular CIS instrument mode). 

There is also a higher energy ($\approx 3$ keV), less dense but hotter population which is strongly modulated by the wave suggesting that some of these energetic ions could be trapped by the large-amplitude wave. The highest count rate of energetic ions on C1 appears during the shock ramp crossing around 15:28:40 UT. This distinct ion population has an energy around 3 keV and should consist of gyrating specularly-reflected protons. A similar but lower count-rate group of ions is observed on the next peak of magnetic intensity. The situation on C3 is slightly different however. There is no distinct energetic population of energetic gyrating ions observed at the "old" shock ramp. These are only seen at the upstream steepened structure, as if it became a new shock front where most of the ion reflection occurs. One notes however that there is little appreciable ion heating between this structure and the old ramp, suggesting that some solar wind plasma is not processed by a full ramp structure but caught in between. 

\begin{figure}
  \includegraphics[width=19pc]{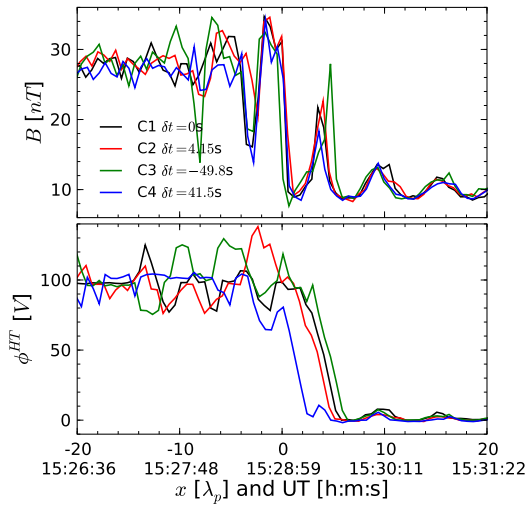}
  \caption{Magnetic field intensity (top) and estimated electric potential (bottom) time-shifted in order to make the magnetic ramps coincide. The downstream cross-shock potential values are very similar on all four spacecraft. However the location of main potential jump seems to move upstream of the main magnetic ramp.}
  \label{fig:cspot}
\end{figure}

Specular ion reflection being associated to the cross-shock potential at least part of the potential jump must occur at the new front. One of the most reliable ways to estimate the electric potential across a shock is to use electron data combined with Liouville mapping (\citep{LSF07JGR} and references therein). Based on assumptions of conservation of electron energy in the de Hoffmann-Teller frame and first adiabatic invariant, this technique relates the changes in electron velocity distributions between two locations to the unknown potential difference. A variant of this technique is used which takes into account the field maxima in between the two locations \citep{LSF07JGR}.

The estimated cross-shock potentials on all four-spacecraft in the de Hoffmann-Teller frame are shown in fig.\ \ref{fig:cspot}, where time-series have been shifted in order to make the main magnetic ramps coincide. Their downstream values are remarkably similar on all four spacecraft despite the shock non-stationarity. However the location of the main potential jump with respect to the main magnetic ramp seems to differ from spacecraft to spacecraft. During the C4 crossing, when the new ramp was just starting to form, the potential jump occurs at the magnetic ramp. It is then observed further upstream on C1 and C2 crossings, and near the newly formed ramp on C3.

\subsection{Large-amplitude downstream perturbations and indications of a reformation cycle}

\begin{figure*}
  \includegraphics[width=0.7\textwidth]{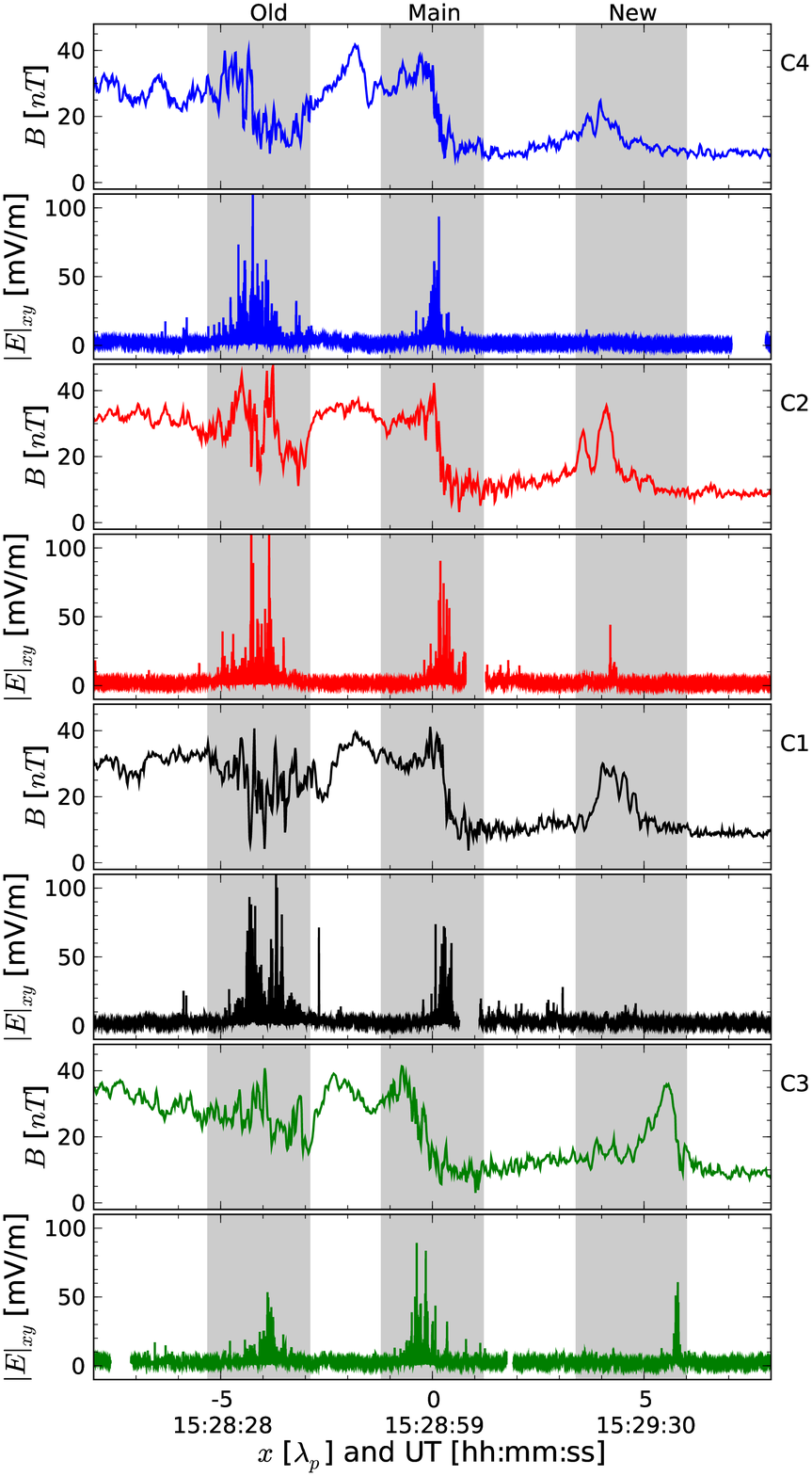}
  \caption{Time-shifted high-resolution magnetic field intensity and electric field ($|E_{xy}|=(E_x^2+E_y^2)^{1/2}$). Spacecraft are ordered in the order by which they cross the shock. Besides the growth of the upstream pulse which becomes a new ramp, data show strong perturbations approximately one wavelength downstream of the main ramp. This can be interpreted as remnants of an old shock front from a previous reformation cycle.}
  \label{fig:cycle}
\end{figure*}

Large perturbations are also found downstream of the main shock ramp as shown in fig. \ref{fig:cycle}, localised at approximately the same distance from the main ramp as the upstream pulse. On all four spacecraft a depression in the mean magnetic field intensity is observed. Large magnetic field and electric field fluctuations are present within this depression. The amplitudes of the electric field fluctuations are comparable to or even larger than at the main ramp. The depth of the depression and amplitude of the fluctuations are also decreasing in time, being the lowest on C3 which is the last to cross the shock. The decrease occurs on time scales comparable to the growth of the upstream pulse. These perturbations can be interpreted as remnants of an older shock ramp which decays in time, showing that the previously described formation of a new ramp is not an isolated event but part of a quasi-periodic reformation cycle. Based on the growth rate of the upstream pulse and the decay of downstream perturbations, the reformation period can be estimated as several tens of upstream proton gyroperiods, and could be a few periods of the upstream low-frequency wave. This is significantly longer than the period of variations of the downstream ion populations ($\sim 2 f_{pu}^{-1}$) observed by \citet{TGB90JGRa}, although the shocks studied by these authors have slightly different parameters and in particular higher Mach number.

\section{Summary and discussion}

The shock analysed in this paper is oblique ($\theta_{Bn}\approx 45^{\circ}$), moderate Mach number ($M_A\approx 5.5$) and low-$\beta$ ($\beta_i\approx 0.1$).

Upstream of the shock a long-wavelength ($\lambda\approx 62\lambda_p$), low-frequency ($f\approx 0.07f_p$ in the plasma frame) and right-hand polarized in the plasma frame quasi-monochromatic wave propagates against the solar wind flow approximately parallel to the upstream magnetic field ($\theta_{kB}\approx 24^{\circ}$). Its properties are consistent with a magnetosonic-like wave excited by a weak ($n_b\ll n$) and cool ($v_{Tb}<v_b$) ion beam backstreaming from the shock through an electromagnetic ion/ion right-hand resonant instability \citep{Gar91SSR}. Such an instability has its maximum growth rate for field-aligned wavevectors and obeys the linear dispersion relation in the cold plasma approximation $\omega = k_{\parallel}v_b - \Omega_p$ (real part). In the low frequency limit $\omega\ll \Omega_p$ one therefore has $k_{\parallel}\approx\Omega_p/v_b$. Assuming that the energetic ion population seen in fig.\ \ref{fig:hia_spectro} ($\sim$3 keV in the spacecraft frame) corresponds to the backstreaming ions yields an estimate for the phase velocity $\omega/k_{\parallel}\approx (\omega/\Omega_p)v_b\approx 80$ km/s in good agreement with the timing results, but the wavelength is found to be $\lambda\approx 8300$ km which is larger than the observed wavelength of 5900 km. It is therefore likely that the 3 keV population (which actually spreads from 2 to 10 keV) is modulated by the wave but not its source. Instead the wave may have been generated further upstream by a beam of ions coming from another part of the shock. Indeed the wave properties described in this paragraph are also similar to those of ULF waves in the ion foreshock \citep{HR83JGR,EBD02GRL}, waves which are generally not observed simultaneously to the field-aligned beams but in association to diffuse-ions or gyrating beams in cyclotron resonance with the wave \citep{MML03PSS}. The observed high energy ion population modulated by the wave is therefore more likely to be either a gyrating beam of reflected ions which managed to escape from the shock or gyrating ions trapped by the wave. 

\begin{figure}
  \includegraphics[width=12pc]{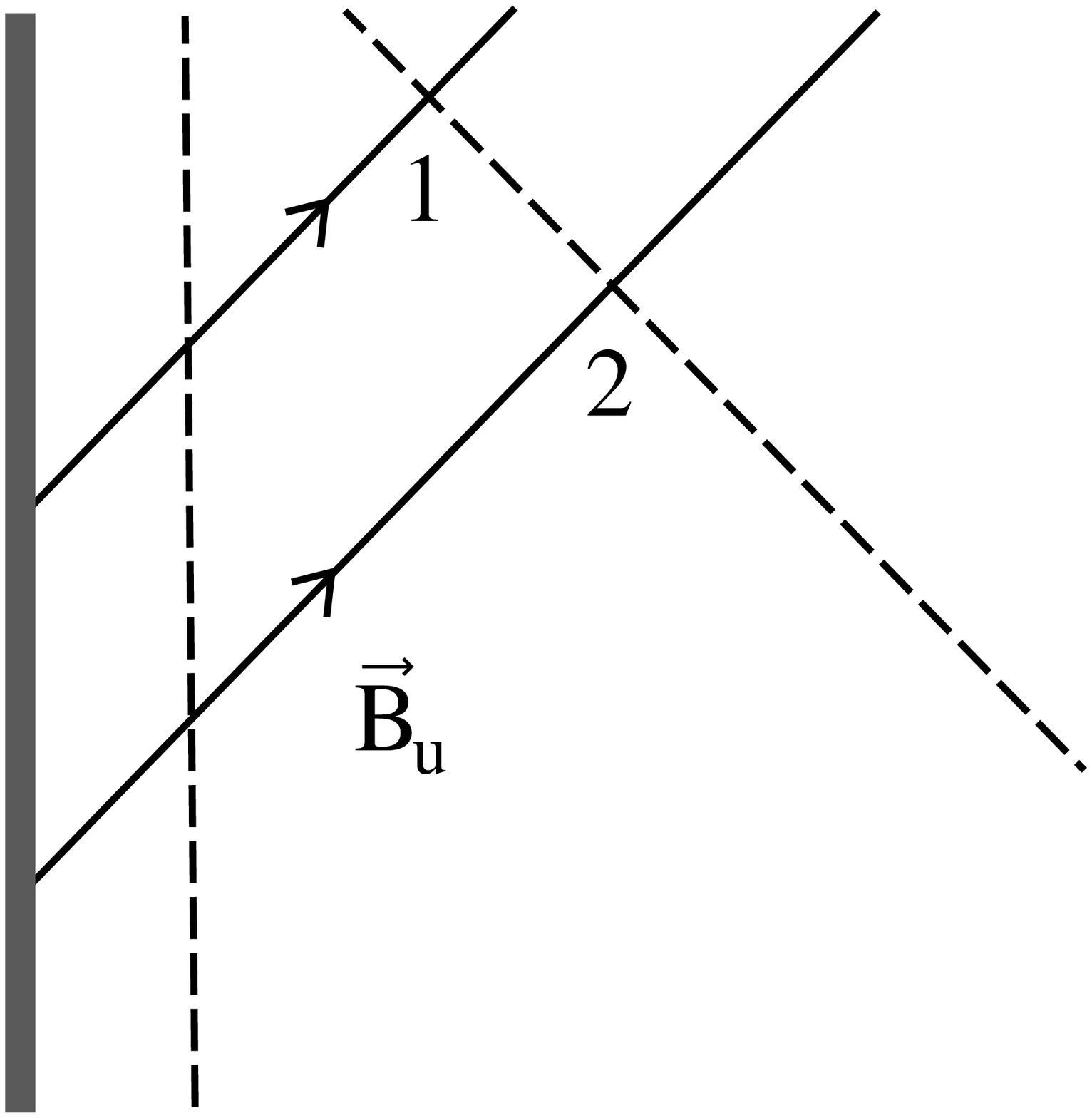}
  \caption{Illustration of the wave refraction near the shock. The wavefronts (dashed lines) become nearly parallel to the shock surface (thick grey line), which requires a lower phase velocity in the shock frame at location 1 (near the shock) than at 2 (further upstream).}
  \label{fig:alignment}
\end{figure}

Due to its small phase velocity the wave is swept back by the solar wind, and closer to the shock the wave properties change significantly. The wavevector aligns itself with the shock normal, which could be due to an increase of refraction index in the foot region. A similar alignment was found in the 2D hybrid simulations of \citet{SFK93JGR}, which they attributed to an increase in density of the diffuse ions near the shock. This alignment makes the perturbations of the shock due to the wave more homogeneous along the shock surface. It is this greater deviation from parallel propagation which allows the wave to become more compressional and steepen, an effect which was conjectured by \citet{HKT87JGR} to happen as ULF waves are convected into areas of the foreshock with higher densities of diffuse ions (see also \citep{BCO06JGR,EBL05JGR}). Besides, the wave has all the appearance of phase-standing with respect to the shock. From a velocity identical to that of the shock in this region, the wavelength is found to be 10 times shorter than further upstream. This slowing down of the wave in the shock frame is entirely consistent with its wavevector alignment with the shock normal, as illustrated by fig.\ \ref{fig:alignment}.

In addition to refractive effects an instability could be operating near the shock front, partly accounting for the changes in wave properties and its strong amplification. This could involve, as dicussed by \citet{WTO90JGR}, a relatively dense and cold beam of specularly reflected ions observed near the shock front during the reformation process in the simulations, or an instability localized at the shock front resulting from the interaction between upstream and some of downstream heated ions. This "interface instability" generates waves on the magnetosonic branch with wavelength intermediate between whistlers and ULF waves. For $M_A=5$ the maximum growth rate corresponds to $k\lambda_p\approx 0.5$ or $\lambda\approx4\pi\lambda_p$. This wavelength ($\lambda\approx 5\lambda_p$) estimated near the shock is smaller, but still in the range where the growth rate is close to its maximum value. The instability is weakly dependant on $\theta_{Bn}$ and tends to have its wavector aligned with the shock normal. However the exact mechanism(s) operating on the wave near the shock front is (are) certainly more complicated since we are dealing with very large-amplitude (nonlinear) waves and a highly inhomogeneous situation. One can however note that because the wavelength near the shock is about 6 times the theoretical wavelength of the phase-standing whistler the mechanism leading to reformation is distinct from the whistler-mediated reformation discussed for oblique and quasi-perpendicular shocks by \citet{BW72JGR,KLS02POP,SB07POP}.

Due to the strong amplification of the wave near the shock, its last cycle before the shock acquires an amplitude comparable to that of the shock itself. This pulsation grows in time, steepens and in the process it emits whistlers, reflects more and more ions and contributes more significantly to the cross-shock potential jump. Reasons for this growth and steepening can be the change in refractive index, a localized instability near the shock front such as the interface instability or nonlinear dynamics of the pulse and its interaction with the shock. Furthermore the amplification mechanism may not operate steadily in time. The negative dips in $B_L$ seen only on C1 and C2 during the intermediate growth stage of the pulse may indeed be the signature of a transient dense ion beam reflected from the shock at this stage of the shock reformation (e.g. as in the simulations of \citet{TWO90JGR}). At least this indicates an unusual and transient processing of ions in between the main ramp and the pulse during the pulse growth.

Whatever mechanism operates here, these observations give strong evidence that the wave cycle standing nearest to the shock is forming a new shock front on a time scale $\sim 35f_{gpu}^{-1}$ or less, comparable to that observed in hybrid simulations of quasi-parallel shocks (e.g. \citep{Bur89GRL,TWO90JGR,SFK93JGR}). The reformation scale (distance between the shock front and pulsation) is $\sim 5\lambda_p$, also in good agreement with simulation results. Assuming that the real shock front thus jumps forward such a distance in a time of $35f_{gpu}^{-1}$, its velocity is found to be only a few km/s larger than that of the "former" shock front in the solar wind frame, which does not induces a significant change in effective Mach number. At approximately the same distance from the main shock ramp but on the downstream side there are magnetic and electric field fluctuations with amplitudes comparable to those at the main ramp. These can be naturally interpreted as remnants of an older ramp which decays on a time scale of at least a few wave periods, which is comparable to the one of the new front formation. These remnants show that the observed reformation is not an isolated event but part of a quasi-periodic reformation cycle.

The reformation process appears quite coherent along the shock surface at the spacecraft separation scale, $\sim 10\lambda_p$, thanks to the wavevector alignment with the shock normal. Some differences in the pulse shape between C1 and C2 measurements are nevertheless observed. The cause of this inhomogeneity could be the finite transverse correlation length of the upstream ULF wave, which is typically of the order of 1$R_E$ according to \citet{LR90JGR} or between 8 and 18 $R_E$ according to \citet{AHL05JGR}. This is however a bit large to account for the inhomogeneity, even when projected on the shock front. It is nevertheless possible that the ULF waves are unstable to transverse perturbations above some threshold, which would reduce their perpendicular correlation length as their amplitude grows. Another reason could be fluctuations or ripples of the shock surface and the inhomogeneity of the ion reflection/wave amplification process. The scale is indeed more consistent with the ripple wavelength (a few tens of $\lambda_p$) estimated by \citet{MBH06JGR} at the quasi-perpendicular bow shock. In any case this inhomogeneity along the shock front is small since the differences between C1 and C2 observations are much smaller than differences with the other two spacecraft. To a first approximation the reformation thus occurs coherently along the shock front on scales of at least $10\lambda_p$.

\section{Conclusion}

The global picture is thus in general agreement with the scenario of quasi-parallel shock reformation proposed by \citet{Bur89GRL}: as some of the specularly reflected ions escape far upstream they generate waves on the magnetosonic branch which are then convected back into the shock and destabilize it. The major difference between the 1d simulations which inspired this scenario and the observations lies in the changes in the wave properties as it nears the shock, where wave refraction, an additional localized instability mechanism such as the interface instability and probably nonlinear phenomena participate in deflecting, slowing down and amplifying the wave cycle which ultimately forms a new shock front. The wave refraction results in a more coherent reformation along the shock surface. On a longer time-scale, this reformation process has no reason to stop unless there is a major change in the upstream conditions and may be repeated quasi-periodically. In this picture the elementary reformation cycle consists of a shock front and wave locally receding at the same speed while the pulse nearest to the "old" shock grows to actually become the "new" shock front upstream of the previous one, the ramp effectively making a jump forward. The old shock slowly vanishes and a new pulse grows upstream of the new one, repeating the reformation cycle.

Finally one may list two key elements which conspire to create this particular shock structure and dynamics lending themselves unusually well to analysis.

First, there is the shock angle $\theta_{Bn}\approx 45^{\circ}$. It means that relatively small changes in this angle have important consequence on the kinematics of the reflected ions, which either cross the shock after half a gyration and thermalise downstream as in quasi-perpendicular shocks or escape upstream and feed ULF waves as in quasi-parallel shocks. Since on average very few of the reflected ions are expected to escape upstream, this angle may be the reason why there is such a large gradient in the reflected ion density near the shock front and therefore why the wave undergoes such large changes in this region. In some sense this leads to a simpler and compacted version of lower-angle shocks. For instance the growing pulse shares some characteristics with SLAMS \citep{SB91GRL}. Indeed evidence suggests that SLAMS grow very rapidly, picking up energy from shock-produced energetic ions, and very possibly only in close proximity to the shock \citep{LHD08JGR}. In addition, they are well-defined entities that propagate from upstream into the nominal shock and give rise to the reformation of the local shock transition. As such, they are
manifestations of the dynamic, structured nature of collisionless shocks under oblique geometries. However their growth mechanism differs from that of the pulse described in this paper, in that they are found in more turbulent ULF wave fields and grow off of gradients in diffuse ion populations rather than the quite mono-energetic ions observed here.

Second, there is the low $\beta_i$. A consequence of this is that specularly reflected ions form a relatively cold beam well distinct from the background ions, helping to shape unstable ion distributions. But the low $\beta$ also makes the specular reflection itself more delicate by making it more difficult for the shock to adjust the electric potential in order to reflect the right amount of ions required for the dissipation of the incoming solar wind kinetic energy. In the zero temperature limit specular reflection even results in a "all or nothing" situation which clearly cannot be steadily sustained, similarly to the 100\% ion reflection or transmission occuring in the high Mach number perpendicular shock simulations of \cite{Que85PRL}. The excess of reflected ions near the shock front is again a likely possibility to explain the wave changes in this region, leading to the formation of a new shock front. 

These two factors show that the shock is in a situation where small changes in $\theta_{Bn}$ or electric potential have major consequences, making the shock unstable and poorly adaptable to changes in the incoming plasma. As it is perturbed major variations in reflected ions strongly modify the energy flow and structure of the shock. The dynamical modifications of the shock scales and structure may then be viewed as a first order attempt by the shock to correct an inappropriate reflection of ions while trying to maintain on average the asymptotic downstream state.

\begin{acknowledgments}
The authors thank M. Fujimoto, S. P. Gary, T. Horbury and I. Shinohara for fruitful discussions.
Parts of this work were supported by a grant to Imperial College from the UK Science and Technology Facilities Council. BL also acknowledges partial support from DOE grant DE-FG02-07ER54941.
\end{acknowledgments}

\end{document}